\documentclass[lettersize,journal]{IEEEtran}

\usepackage{amsthm}
\usepackage{amsmath}
\usepackage{amsfonts}
\usepackage{amssymb}
\usepackage{mathrsfs}
\usepackage{bm}
\usepackage{graphicx}
\usepackage{subfigure}
\usepackage{stfloats}
\usepackage{booktabs}
\usepackage{hyperref}
\usepackage{cite}
\usepackage{cases}
\usepackage{tabularx}
\usepackage{textcomp}
\usepackage{float}
\usepackage[ruled,linesnumbered,vlined]{algorithm2e}

\usepackage{textcomp}

\begin{document}
\title{Energy-Efficient Hybrid Data Computation via Coordinated AirComp and Edge Offloading}
\author{ Yudan Jiang, Xiao Tang, Jinxin Liu, Qinghe Du, Dusit Niyato, and Zhu Han
\thanks{Y. Jiang is with Shenzhen Research Institute of Northwestern Polytechnical University, Shenzhen 518057, China, and also with the School of Electronics and Information, Northwestern Polytechnical University, Xi'an 710072, China.}
\thanks{X. Tang is with the School of Information and Communication Engineering, Xi'an Jiaotong University, Xi'an 710049, China, and also with Shenzhen Research Institute of Northwestern Polytechnical University, Shenzhen 518057, China.. (e-mail: tangxiao@xjtu.edu.cn)}
\thanks{J. Liu is with the School of Mechanical Engineering, Xi’an Jiaotong University, Xi’an 710049, China.}% 
\thanks{Q. Du is with the School of Information and Communication Engineering, Xi'an Jiaotong University, Xi'an 710049, China.}
\thanks{D. Niyato with the College of Computing and Data Science, Nanyang Technological University, Singapore.}
\thanks{Z. Han is with the Department of Electrical and Computer Engineering, University of Houston, Houston 77004, USA.}
}

\maketitle

\begin{abstract}
The development of 6G networks brings an increasing variety of data services, which motivates the hybrid computation paradigm that coordinates the over-the-air computation (AirComp) and edge computing for diverse and effective data processing. In this paper, we address this emerging issue of hybrid data computation from an energy-efficiency perspective, where the coexistence of both types induces resource competition and interference, and thus complicates the network management. Accordingly, we formulate the problem to minimize the overall energy consumption including the data transmission and computation, subject to the offloading capacity and aggregation accuracy. We then propose a block coordinate descent framework that decomposes and solves the subproblems including the user scheduling, power control, and transceiver scaling, which are then iterated towards a coordinated hybrid computation solution. Simulation results confirm that our coordinated approach achieves significant energy savings compared to baseline strategies, demonstrating its effectiveness in creating a well-coordinated and sustainable hybrid computing environment.
\end{abstract}

\begin{IEEEkeywords}
Energy minimization, edge computing, over-the-air computation, integrated communication and computation.
\end{IEEEkeywords}

\section{Introduction}
\IEEEPARstart{T}{he} deep integration of the communication and computation is the key paradigm of the 6G technology, enabling the network with simultaneous data transmission and processing~\cite{05}. The revolution is important for the applications such as the autonomous driving and environmental sensing. Meanwhile, the rapid growth of computational demands of terminals and edge nodes in the 6G era has induced a rapid climb in energy consumption~\cite{02}. Consequently, it is difficult for traditional separated communication and computation frameworks to balance the performance and energy efficiency. The integration of the communication and computation motivates the seek for new paths to save energy by decreasing the excess data transfer overhead and raising the overall energy efficiency performance of the system through co-design. Recent studies have explored the integration of sensing, communication, and computation in next-generation wireless networks. For example, a systematic framework for Integrated Sensing, Communication, and Computation (ISCC) oriented toward 6G-enabled ubiquitous intelligence was developed in \cite{extra001}. Therefore, designing energy-efficient integrated communication and computation systems has become essential for the sustainable development of 6G networks~\cite{zhr}.

Towards the vision of efficient data computation, edge computing has been a promising solution to alleviate network congestion and reduce the transmission distance, lowering the system energy~\cite{17}. Moreover, over-the-air computation (AirComp) is an emerging technology to enable more efficient data processing in the scenarios such as the massive number of the nodes need to be processing in time~\cite{08, 09}. Existing work mainly addresses these two related techniques separately, such as joint optimization of communication and computation resources~\cite{11, 20n}, task scheduling and resource allocation~\cite{12, 21n}, energy-aware system design~\cite{12} and robustness against channel impairments~\cite{01}. 
 
However, in the practical applications such as smart sensing, environmental monitoring and industrial IoT, heterogeneous computing demand tasks usually coexist in the same scenario. This data diversity emphasized the relevance of supporting the hybrid computing paradigm instead of relying on one single manner. To meet this challenge, the coordination of the edge computing and AirComp, noted as hybrid computing has been a promising solution to meet the request of heterogeneous services~\cite{13,14}. Particularly, for the energy-constrained devices, the network can through dynamically seek the optimal computation strategy to degrade the node power consumption.

Despite the potential of coordinated AirComp and edge processing, the implementations of such hybrid computing framework can be rather challenging. These challenges present an urgent need  to design a new paradigm that can effectively coordinate the coexistence of these two ways of data compuation. To bridge this gap, we investigate a hybrid computing system that integrates edge computing with AirComp with energy-efficiency-oriented operations. In particular, the main contributions can be summarized as follows:
\begin{itemize}
	\item We propose a novel hybrid wireless computing paradigm with concurrent AirComp for real-time function aggregation and edge offloading for data-intensive tasks, sharing the network resources for effective data processing.
	\item We formulate a problem to minimize the overall energy consumption of the hybrid data computation by jointly optimizing the user scheduling, power control, and transceiver scaling, while considering the offloading capacity and AirComp accuracy constraints. 
	\item We employ the block coordinate descent (BCD) framework to solve the complex non-convex energy optimization problem, which decouples the variables into a series of convex subproblems and further updates iteratively towards the solution.
\end{itemize}

\section{System Model}

\begin{figure}
	\centering
	\includegraphics[width=0.45\textwidth]{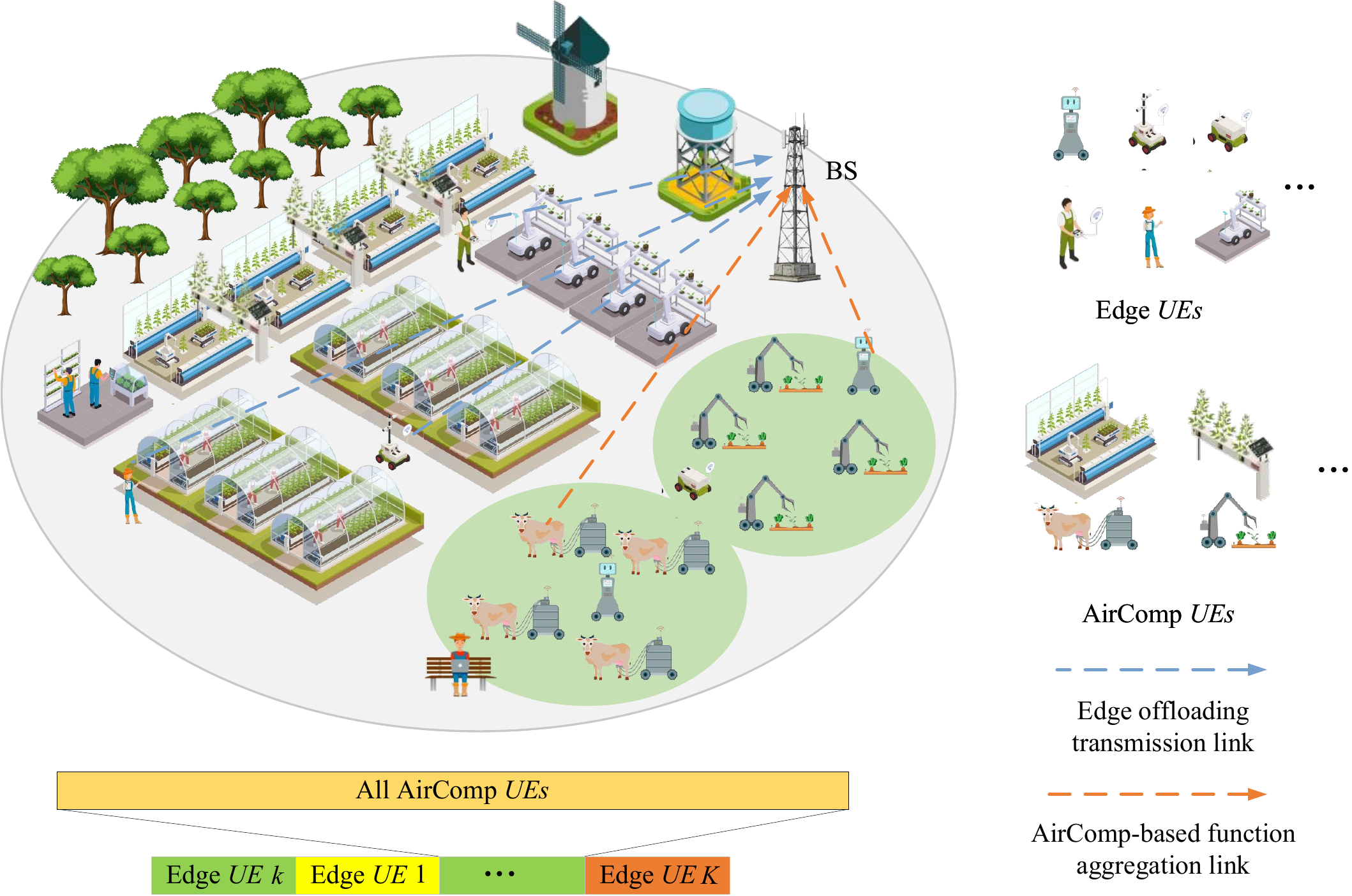}
	\caption{System model.}
	\label{sys}
\end{figure}

As shown in Fig.~\ref{sys}, we consider a hybrid wireless computing system that integrates both AirComp and edge computing to meet different task requirements. The system consists two types of user equipment (UE) and a base
station (BS). Among the UEs, some are assigned to perform AirComp-based functional aggregation, forming the set $\mathcal{J}$, while others are designated for edge task offloading, forming the set $\mathcal{K}$. These two sets are disjoint, $\mathcal{J} \cap \mathcal{K} = \emptyset$, and their union forms the complete set of UEs in the system, denoted by $\mathcal{S} = \mathcal{J} \cup \mathcal{K}$. It is assumed that the locations of the ground UEs and the BS are fixed, in order to highlight the performance limits of the proposed hybrid computation and resource allocation framework while keeping the problem tractable. To efficiently coordinate communication and computation, we adopt a time-slotted operation model. The entire time horizon $T$ is evenly divided into $I$ slots, denoted by $\mathcal{I} = \left\{ 1, 2, \ldots, I \right\}$, each with duration $\frac{T}{I}$. In each time slot-$i$, all AirComp UEs in $\mathcal{J}$ simultaneously transmit their signals for AirComp, while only one edge UE from $\mathcal{K}$ is scheduled to offload its computation task to the BS. The BS simultaneously receives the superimposed AirComp signal and the edge task offloading data.

Given the stationary deployment of both the BS and UEs in our scenario, the wireless propagation environment is characterized by a dominant line-of-sight (LoS) path and relatively relatively weak non-line-of-sight (NLoS) components. Hence, we model the wireless channel as a Rician fading channel, as shown below 
\begin{equation}
	\begin{aligned}	
		h_{s}(i) = \sqrt{\frac{\beta_0} {d_{s}^{2}}}\left(\sqrt{\frac{\kappa(i)}{\kappa(i) + 1}}h^{\text{LoS}}_{s}(i) + \sqrt{\frac{1}{\kappa(i) + 1}}h^{\text{NLoS}}_{s}(i)\right),&\\ \quad \forall s \in \mathcal{S}, \quad \forall i \in \mathcal{I},&
	\end{aligned}
\end{equation}
where $d_{s}$ is the distance between UE $s$ and the BS, and $\beta_0$ represents the path loss at the reference distance, and $\kappa(i)$ represents the $\text{Rician}$ factor for each timeslot channel. $h^{\text{LoS}}_{s}(i) $ and $h^{\text{NLoS}}_{s}(i) $ represent the LoS and NLoS component respectively.

\subsection{Signal Model}
At each time slot, AirComp UE $j$ transmits pre-processed signals $s_j$ to the BS for aggregation. Meanwhile, one selected edge UE $k$ offloads a data packet $x_k$ to the BS for edge computation. The AirComp UEs and edge UEs transmit pre-processed data associated with the current slot rather than repeatedly transmitting identical data. Without loss of generality, we assume that the signals are independent among different sensors with zero mean and normalized variance. The edge computing offloading activity is indicated by a binary variable $\alpha_k(i)$,  
\begin{equation} \label{binary} % equ1 \ref{equ1} 
	\left\{ \begin{aligned}
		&\alpha_k\left(i\right) \in \left\{ {0,1} \right\}, &&\quad \forall k \in \mathcal{K},\:\forall i \in \mathcal{I},\\
		&\sum\limits_{k \in \mathcal{K}} {\alpha_k\left(i\right)} = 1, &&\quad \forall i \in \mathcal{I},
	\end{aligned} \right.
\end{equation}
where $\alpha_k(i) = 1$, if edge UE $k$ is selected in time slot-$i$, and $\alpha_k(i) = 0$, otherwise. Thus, the aggregated signal received at the BS is
$
	y\left( i \right) =  \sum\limits_{j \in \mathcal{J}} h_j(i) b_j(i) s_{j} +  \sum\limits_{k \in \mathcal{K}}\alpha_k(i)h_k(i) \sqrt{p_k(i)} x_{k} +  n(i), \quad \forall i \in \mathcal{I},
$
where $h_j(i), j \in \mathcal{J}$, is the channel between BS and the AirComp UEs and $h_k(i), k \in \mathcal{K}$, is the channel between BS and the edge UEs. In this paper, we assume the channel state information (CSI) is perfectly known, which can be obtained by pilot-based estimation~\cite{pilot}. $n(i) $ denotes the receiver additive white Gaussian noise with the background noise power $\sigma_0^2$, and $b_j(i) $ is the Tx-scaling factor of the AriComp UE $ j $, and $p_k(i)$ is the transmit power of the edge UE $k$. The transmit power constraints for both types of UEs are given as
\begin{equation}\label{p_edge}
	0 \leq  p_k{(i)} \leq P_k^{\text{(E, max)}}, \quad \forall k \in \mathcal{K}, \quad \forall i \in \mathcal{I},
\end{equation}
and
\begin{equation}\label{p_air}
\left| b_j(i) \right|^2 \leq P_j^{\text{(A, max)}},\quad \forall j \in \mathcal{J}, \quad \forall i \in \mathcal{I},
\end{equation}
where $P_k^{\text{(E, max)}}$ and $P_j^{\text{(A, max)}}$ denote the respective power budgets of the edge UE $k$ and the AirComp UE $j$. 

\subsection{AirComp Model}
In the AirComp framework, the BS aims to compute a pre-defined function of the transmitted signals, take the sum as an example due to its linearity and applicability to various learning and sensing tasks. Therefore, the target function in time slot-$i$ is defined as
$	\hat{y}(i) = \sum_{j \in \mathcal{J}} s_{j}, \quad \forall i \in \mathcal{I}.$
To compute the desired results, which is theoretically known to be the target aggregation result, we need to recover it from the received superimposed signal at the BS. To this end, we introduce the Rx-scaling factor $\eta\left( i \right) $ at time slot-$i$. Accordingly, the 
	$\tilde{y}\left( i \right) = {\eta\left( i \right)}y\left( i \right),  \quad \forall i \in \mathcal{I},
$
where $y\left( i \right)$ is the aggregated signal received at the BS. To evaluate the accuracy in the AirComp process, we define the computation distortion using the mean squared error (MSE), defined for each time slot-$i$ as
\begin{equation}\label{11MSE}	
	\mathsf{MSE}\left( i \right) = \mathbb{E} \left\{ \left| \tilde{y}\left( i \right) - \hat{y}\left( i \right) \right|^2 \right\}, \quad \forall i \in \mathcal{I}.
\end{equation}
By substituting $y\left( i \right)$ and $\hat{y}\left( i \right) $ into~(\ref{11MSE}), the expression of the MSE becomes
\begin{equation}
	\begin{aligned}	
		\mathsf{MSE}\left( i \right) = & \sum_{j \in \mathcal{J}} \left| {{\eta(i)}} b_j(i) h_j(i) - 1 \right|^2 + \left| {\eta(i)} \right|^2 \sigma_0^2 \\
		& + \sum\limits_{k \in \mathcal{K}}\left|{\eta(i)}\alpha_k(i)\sqrt{p_k(i)} h_k(i)\right|^2, \quad \forall i \in \mathcal{I}.
	\end{aligned}
\end{equation}
To ensure reliable AirComp computation under hybrid transmission, the system MSE is constrained below a predefined threshold $\zeta$, i.e., 
\begin{equation}\label{MSE}
    \mathsf{MSE}\left( i \right) \leq \zeta, \quad \forall i \in \mathcal{I}.
\end{equation}
The constraint in ~(\ref{MSE}) captures aggregation distortion, receiver noise, and interference from edge UEs, which ensures that the computation result remains reliable under hybrid transmission.

\subsection{Edge Computing Model}

We consider that each edge UE $k \in \mathcal{K}$ generates a total of $D_k$ bits of data over the time horizon $T$, which needs to be offloaded to the BS for remote processing. To this end, the total data is distributed across the $I$ time slots, with $l_k(i)$ denoting the amount of data offloaded by UE $k$ in slot $i$. Accordingly, the total offloaded data over all time slots should satisfy the following requirement
\begin{equation}\label{data_offload}
	\sum\limits_{i \in \mathcal{I}} l_k\left( i \right) \geq D_k, \quad \forall k \in \mathcal{K}.
\end{equation}
To ensure the offloaded tasks can be processed by the BS, the computing workload in each slot must not exceed its processing capacity. Let $c_0$ denotes the required CPU cycles per bit, and let $f^{\text{(max)}}$ be the maximum computation frequency of BS. Then, the edge computation at slot-$i$ should satisfy the following constraint
\begin{equation}\label{data_ability}
	c_0 \sum_{k \in \mathcal{K}}\alpha_k(i)l_k\left( i \right) \leq \frac{T}{I} f^{\text{(max)}}, \quad \forall i \in \mathcal{I}.
\end{equation}
During the slot-$i$, the data offloaded for edge computing at the edge UE $k$ is restricted by the transmission constraint
\begin{equation}\label{trans_edge}
	\begin{aligned}
	\frac{T}{I} B\log_2\left( 1 + \frac{p_k(i) \left|h_k(i)\right|^2}{\sum\limits_{j \in \mathcal{J}} \left|h_j(i) b_j(i)\right|^2 + \sigma_0^2} \right) \geq l_k\left( i \right),&\\ \quad \forall k \in \mathcal{K}, \quad \forall i \in \mathcal{I}.&
	\end{aligned}
\end{equation}
where $B$ is the communication bandwidth.

\subsection{Energy Model}

Based on the model presented above, the energy of the network consists of three parts, i.e., the transmission energy of the edge UEs for offloading computation, the transmission energy of the AirComp UEs, and the computation energy of BS. Specifically, the transmission energy of the edge computing UE $k$ at the slot-$i$ is
$
	E^{\text{(E, tran)}} = \sum_{i \in \mathcal{I}} \sum_{k \in \mathcal{K}} \alpha_k(i)p_k\left( i \right)\frac{T}{I}.
$
The transmission energy of the AirComp UE $k$ at the slot-$i$ is
$
	E^{\text{(A, tran)}} = \sum_{i \in \mathcal{I}}\sum_{j \in \mathcal{J}}  \left| b_j(i) \right|^2 \frac{T}{I}.
$
The computation energy of the BS is expressed as
$
	E^{\text{(comp)}} = \sum_{i \in \mathcal{I}} \frac{\gamma {\left( c_0 \sum\limits_{k \in \mathcal{K}}\alpha_k(i)l_k\left( i \right) \right)^3}}{\left(\frac{T}{I}\right)^2},
$
where $\gamma$ is the effective capacitance coefficient of the BS. The equation references the cubic relationship between the energy consumption of edge devices and computational load. Since AirComp performs functional computation directly over the air, its computation energy is negligible and thus omitted from the total energy cost. Therefore the total energy cost can be given as $E^{\text{(sum)}} = E^{\text{(E, tran)}} + E^{\text{(A, tran)}} + E^{\text{(comp)}}$.

\section{Proposed Solution}
In this section, we formulate the problem into minimizing the energy consumption. By analyzing the complex problem we decomposed it into several subproblems. The algorithm is proposed to achieve energy-efficient integrated communication and computation in hybrid wireless computing system.
\subsection{Problem Formulation and Decomposition}
For our considered model, both types of UEs compete over limited bandwidth and power resources, leading to tightly coupled transmission and computation costs. To improve energy efficiency, it is crucial to jointly manage the transmission and computation. In this context, we formulate an energy minimization problem that jointly optimizes the transmission strategies of edge UEs and AirComp UEs, as well as the data offloading decisions of edge computing UEs. Although the primary objective is energy minimization, the proposed model is able to naturally incorporate both transmission and computation delays, thereby enabling latency-aware optimization under practical application requirements.
The optimization problem is given as follows
\begin{IEEEeqnarray}{CL} \label{eq:problem_equiv}
	\IEEEyesnumber \IEEEnosubnumber*
	\min_{\substack{ \left\{ {\alpha_k(i),l_k(i),p_k(i)} \right\}_{\forall k \in \mathcal{K}, \forall i \in \mathcal{I}} \\ \left\{ {b_j(i),\eta(i)} \right\}_{ \forall j \in \mathcal{J}, \forall i \in \mathcal{I} } }} & E^{\text{(sum)}} \\
	{\rm{s.t.}} 
	& (\ref{binary}), (\ref{p_edge}), (\ref{p_air}),  (\ref{MSE}), (\ref{data_offload}), (\ref{data_ability}), (\ref{trans_edge}). \nonumber \IEEEeqnarraynumspace
\end{IEEEeqnarray}
It can be seen that the formulated problem exhibits multi-fold layers of complexity. First, the variables are coupled, jointly affecting both the energy consumption and system constraints. Second, the transmission constraint involves a non-convex logarithmic SINR expression, where the SINR term is quadratic, further complicating the feasible region. Moreover, the time-coupled nature of offloading and computation across multiple slots makes direct optimization intractable.

To facilitate tractable optimization, we decompose the original problem into three subproblems. Each subproblem is addressed individually, and their solutions are integrated using the block coordinate descent (BCD) framework~\cite{BCD}. The decomposition of problem (10) is motivated by the strong coupling among variables and their distinct physical functionalities. By grouping variables with similar roles into separate blocks and adopting the BCD framework, the original non-convex problem is transformed into several tractable convex subproblems, effectively alleviating variable coupling and enabling efficient optimization. The first is the Rx-scaling factor at the BS and the amount of data offloaded by each edge UE in each time slot. Therefore, the first problem concerns time-slotted offloading data amount and Rx-scaling factor, given as 
\begin{IEEEeqnarray}{CL} \label{eq:problem_1}
	\IEEEyesnumber \IEEEnosubnumber*
	\min_{\substack{  \left\{ \eta(i),l_k(i) \right\}_{ \forall k \in \mathcal{K}, \forall i \in \mathcal{I} }}}  & E^{\text{(comp)}} \\
	{\rm{s.t.}} &  (\ref{MSE}), (\ref{data_offload}), (\ref{data_ability}), (\ref{trans_edge}). \nonumber
	\IEEEeqnarraynumspace
\end{IEEEeqnarray}
The centralized optimization of AirComp transmit scaling factors is efficient because these variables are strongly coupled through global CSI, MSE constraints, and system-level energy tradeoffs, which are difficult to handle with purely distributed or heuristic methods. With given data offloading amount, the second subproblem is the transmit power allocation for both edge UEs and the AirComp UEs, specified as
\begin{IEEEeqnarray}{CL} \label{eq:problem_2}
	\IEEEyesnumber \IEEEnosubnumber*
	\min_{\substack{  \left\{ b_j(i) \right\}_{ \forall j \in \mathcal{J}, \forall i \in \mathcal{I}}\\ \left\{ {p_k(i)} \right\}_{ \forall k \in \mathcal{K}, \forall i \in \mathcal{I} }}} & E^{\text{(A, tran)}} + E^{\text{(E, tran)}} \\
	{\rm{s.t.}} &  (\ref{p_edge}), (\ref{p_air}), (\ref{MSE}), (\ref{trans_edge}). \nonumber
	\IEEEeqnarraynumspace
\end{IEEEeqnarray}
Finally, given all the information, the problem for which edge UE the BS should schedule for computation offloading in each time slot to minimize the overall energy consumption is cast as
\begin{IEEEeqnarray}{CL} \label{eq:problem_3}
	\IEEEyesnumber \IEEEnosubnumber*
	\min_{\substack{  \left\{ \alpha_k{(i)} \right\}_{ \forall k \in \mathcal{K}, \forall i \in \mathcal{I} } }} & E^{\text{(E, tran)}} + E^{\text{(comp)}} \\
	{\rm{s.t.}} &  (\ref{binary}), (\ref{MSE}), (\ref{data_ability}). \nonumber
	\IEEEeqnarraynumspace
\end{IEEEeqnarray}
It can be observed that the adopted BCD framework is able to decompose the original problem into several subproblems, each of which can be solved exactly and efficiently, leading to an effective approximation to the solutions to the original problem. With the subproblems identified, we next focus on analyzing and solving the decomposed subproblems to obtain the final solution to the original problem.

\subsection{Subproblem Analysis}
For the first subproblem of~(\ref{eq:problem_1}), given that the objective function and all constraints are linear with respect to the decision variables, the problem falls into the class of linear programming, which is known for its tractability and can be efficiently solved using mature optimization toolkits.

For the second subproblem in~(\ref{eq:problem_2}), the problem exhibits linearity in both the objective function and the constraints specified in~(\ref{p_edge}), (\ref{p_air})  and~(\ref{MSE}). However, the transmission constraint in~\eqref{trans_edge} introduces non-convexity due to the coupling between the transmit power $p_k(i)$ and $b_j(i)$ in the SINR term. In this respect, we introduce an auxiliary variable $\psi_j(i)$ to relax the interference term, such that
\begin{equation}\label{slack}
    \psi_j(i) \geq \left|b_j(i)\right|^2, \quad \forall j \in \mathcal{J}, \quad \forall i \in \mathcal{I},
\end{equation}
which induces that
$
		\frac{T}{I} B\log_2\left( 1 + \frac{p_k(i) \left|h_k(i)\right|^2}{\sum\limits_{j \in \mathcal{J}} \left|h_j(i) \psi_j(i)\right| + \sigma_0^2} \right) \geq l_k\left( i \right), \quad \forall k \in \mathcal{K}, \quad \forall i \in \mathcal{I}.
$
Further, we reorganize the terms and transform into its equivalent exponential representation, as shown below
\begin{equation}\label{trans_edge_slack_2}
	\begin{aligned}
		p_k(i) \left|h_k(i)\right|^2 \geq \left( 2^{\frac{l_k\left( i \right) I }{TB}} - 1 \right)\left(\sum\limits_{j \in \mathcal{J}} \left|h_j(i)\right|^2 \psi_j(i) + \sigma_0^2\right),&\\ \quad \forall k \in \mathcal{K}, \quad \forall i \in \mathcal{I}.&
	\end{aligned}
\end{equation}
Evidently, the resulting constraint~(\ref{slack}), together with constraint~(\ref{trans_edge_slack_2}) after the substitution of variables, are both convex. As a result, all the non-convex components in subproblem (13) are eliminated, and subproblem (13) is reformulated as a convex optimization problem with respect to the variables $b_j(i)$, $\psi_j(i)$, and $p_k(i)$. Building upon the preceding analysis, we derive the convex counterpart of the problem in~(\ref{eq:problem_2}) as follows,
\begin{IEEEeqnarray}{CL} \label{eq:problem_2_slack}
	\IEEEyesnumber \IEEEnosubnumber*
	\min_{\substack{  \left\{ b_j(i),\psi_j(i) \right\}_{ \forall j \in \mathcal{J}, \forall i \in \mathcal{I}}\\ \left\{ {p_k(i)} \right\}_{ \forall k \in \mathcal{K}, \forall i \in \mathcal{I} }}} & E^{\text{(A, tran)}} + E^{\text{(E, tran)}} \\
	{\rm{s.t.}} &  (\ref{p_edge}), (\ref{p_air}), (\ref{MSE}), (\ref{slack}), (\ref{trans_edge_slack_2}). \nonumber
	\IEEEeqnarraynumspace
\end{IEEEeqnarray}

For the third subproblem in~(\ref{eq:problem_3}), the optimization variables $\alpha_k{(i)}, \forall k \in \mathcal{K}, \forall i \in \mathcal{I} $ are binary indicators representing whether edge UE $k$ is scheduled in time slot-$i$. Due to the computational complexity associated with integer programming, we relax these binary variables to continuous variables, i.e.,
\begin{equation} \label{binary_slack} % equ1 \ref{equ1} 
	\left\{ \begin{aligned}
		&\alpha_k{(i)} \in \left[0, 1\right], &&\quad \forall k \in \mathcal{K},\:\forall i \in \mathcal{I},\\
		&\sum\limits_{k \in \mathcal{K}} {\alpha_k\left(i\right)} = 1, &&\quad \forall i \in \mathcal{I}.
	\end{aligned} \right.
\end{equation}
Applying the relaxation yields a convex problem, as the objective is convex and all constraints are affine with respect to the variable, expressed as
\begin{IEEEeqnarray}{CL} \label{eq:problem_3_slack}
	\IEEEyesnumber \IEEEnosubnumber*
	\min_{\substack{  \left\{ \alpha_k{(i)} \right\}_{ \forall k \in \mathcal{K}, \forall i \in \mathcal{I} } }} & E^{\text{(E, tran)}} + E^{\text{(comp)}} \\
	{\rm{s.t.}} &  (\ref{MSE}), (\ref{data_ability}),(\ref{binary_slack}). \nonumber
	\IEEEeqnarraynumspace
\end{IEEEeqnarray}
Upon obtaining the continuous solution, in the algorithm implementation, a max-value rounding recovery strategy is adopted to map the continuous solution to scheduling decisions that satisfy the original binary constraints, thereby strictly ensuring the feasibility of the time-slot scheduling constraints.

\subsection{Algorithm Design}
In the preceding discussions, we have analyzed each decomposed subproblem and proposed strategies accordingly. When tackling a particular subproblem, the proposed methods operate under the assumption that the variables associated with the other subproblems are fixed. We denote the optimization variable group $\left\{\left\{ \eta(i),l_k(i) \right\}_{ \forall k \in \mathcal{K}, \forall i \in \mathcal{I} }\right\}$, and $\left\{\left\{ b_j(i) \right\}_{ \forall j \in \mathcal{J}, \forall i \in \mathcal{I}}, \left\{ {p_k(i)} \right\}_{ \forall k \in \mathcal{K}, \forall i \in \mathcal{I} }\right\}$, and $\left\{\left\{ \alpha_k{(i)} \right\}_{ \forall k \in \mathcal{K}, \forall i \in \mathcal{I} }\right\}$ as $\bm{\Omega} $, $\bm{\Theta}$, and $\bm{\Xi}$. The algorithm terminates when the reduction in energy consumption falls below a predefined threshold $\varepsilon_0$. The procedure is summarized in Alg.~\ref{alg}.
\SetKwInput{KwIn}{Initialization}
\begin{algorithm}\label{alg}\small
	\renewcommand{\thealgocf}{1}
	\SetAlgoLined
	\KwIn{randomly select ${\bm{\Omega} _{\rm{0}}}$, ${\bm{\Theta} _{\rm{0}}}$ and ${\bm{\Xi} _0}$ satisfying the constraints in~(\ref{eq:problem_equiv}).}
	\Repeat{$\left| {{E_{0}} - E\left( {{\bm{\Omega} ^ * };{\bm{\Theta} ^ * };{\bm{\Xi} ^ * }} \right)} \right| /\left|{E_{0}}\right|< {\varepsilon _{0}}$}
	{
		Solve problem~(\ref{eq:problem_1}) with given ${\bm{\Omega} _{\rm{0}}}$, ${\bm{\Theta} _{\rm{0}}}$, ${\bm{\Xi} _0}$, and find the optimal denoted as ${\bm{\Omega} ^ * }$;\\
		Solve problem~(\ref{eq:problem_2_slack}) with given ${\bm{\Omega} ^ * }$, ${\bm{\Theta} _{\rm{0}}}$, ${\bm{\Xi} _0}$, and find the optimal denoted as ${\bm{\Theta} ^ * }$;\\
		Solve problem~(\ref{eq:problem_3_slack}) with given ${\bm{\Omega} ^ * }$, ${\bm{\Theta}^*}$, ${\bm{\Xi} _0}$, and find the optimal denoted as ${\bm{\Xi} ^ * }$;\\
		${\bm{\Omega} _{\rm{0}}} \leftarrow {\bm{\Omega}^ *},  {\bm{\Theta} _{\rm{0}}} \leftarrow {\bm{\Theta}^ *}, {\bm{\Xi} _0} \leftarrow {\bm{\Xi}^ *}$;
	}

	\caption{BCD for energy-minimized hybrid computing}
\end{algorithm} 
Before executing the algorithm, the system first performs pilot-assisted estimation to obtain the required channel information. Subsequently, the proposed joint optimization algorithm is carried out based on the estimated channel information. During the algorithm initialization stage, all optimization variables are generated randomly, but the initial values are selected only from the feasible set that satisfies all constraints in problem (11), thereby ensuring the feasibility of the initial solution. As the proposed algorithm follows a BCD framework, the convergence is
	justified in the following aspects. Since each subproblem intends to minimize the energy
	consumption, the overall energy update process is non-increasing during the iterations. Furthermore, the energy minimization problem is well-defined on a compact set and thus is bounded. Furthermore,
	based on the classical convergence results of BCD methods for non-convex optimization
	problems, when the objective function is continuously differentiable and each subproblem is
	solved exactly, the generated sequence converges to a stationary point.

Finally, the complexity of proposed approach is analyzed as follows. Generally, the complexity of the proposed algorithm encompasses the computations for solving the three convex subproblems in each BCD iteration, as well as the total number of iterations required for convergence. The problem in (11) incorporates $I(1 + K)$ decision
	variables, the computation complexity is then $\mathcal{O}\left((I(1 + K))^{3.5}\right)$. Similarly, the problem solving
	of (16) also has complexity of $\mathcal{O}\left((I(2J + K))^{3.5}\right)$, since it has $I(K + J)$
	decision variables and $IJ$ slack variables. The problem in (18) incorporates $IK$ decision
	variables, the computation complexity is then $\mathcal{O}\left((IK)^{3.5}\right)$. A single BCD iteration requires sequentially solving these three subproblems. Therefore, the complexity of a single iteration is the sum of the complexities of the three subproblems, which is given by $\mathcal{O}\left((I(1 + K))^{3.5} + (I(2J + K))^{3.5} + (IK)^{3.5} \right)$. Assuming that the algorithm converges after $M$ iterations, the overall complexity of the proposed algorithm is given by the cumulative complexity across all iterations, which is given by $\mathcal{O}\left(M\left((I(1 + K))^{3.5} + (I(2J + K))^{3.5} + (IK)^{3.5} \right)\right)$.

\section{Simulation Results}
In this section, we present the simulation results to demonstrate the performance of our proposed approach. Within a square area with a side length of 1000 meters, 20 nodes are randomly distributed. The parameters used in the simulation are detailed in Table~\ref{tab}. Specifically, all simulations were conducted using the SDPT3 solver on MATLAB. We consider the following two schemes as performance comparison benchmarks.
\begin{itemize}
	\item \textit{Equal offloading transmission}: 
Edge node $k$ uniformly offloads its pending data volume during the scheduled time slot.
	\item \textit{Channel-inversion power control}\cite{MSE1, MSE2}: Each Aircomp UE $ j $ sets the Tx-scaling factor based on the channel inversion principle, i.e., $b_j(i) = \sqrt{P_j^{\text{(A, max)}}}\frac{\min_{l \in \mathcal{J}} \left|h_l(i)\right|}{\sqrt{\left|h_j(i)\right|^2 + \sigma_0^2 }}\frac{h_j(i)}{\left|h_j(i)\right|}$.
\end{itemize}

\begin{table} \small
	\caption{Simulation parameters}
	\centering
	\begin{tabular}{ cc|cc }
		\toprule
		Notation    & Value & Notation & Value \\
		\midrule
		$T$                &  $200\,\text{s}$         & $f^{(\text{max})}$                &  $6\,\text{GHz}$ \\
		$I$                &  $200$                   & $P_j^{\text{(A, max)}}$      &  $1\,\text{W}$ \\
		$B$                &  $5\,\text{MHz}$                    & $P_k^{\text{(E, max)}}$ &  $1\,\text{W}$ \\
		
		% Energy and motion
		$D_k$              &  $6\,\text{Mbits}$         & $\sigma_0^2$   &  $-120\,\text{dBm}$ \\
		$c_0$                &  $1\times10^{3}\,\text{cycles/bit}$                    & $\gamma$           &  $1\times10^{-27}$ \\
		$\kappa$                &  $15$                    & $\beta_0$           &  $-60\,\text{dBm}$ \\
		\bottomrule
	\end{tabular}\label{tab}
\end{table}

\begin{figure*}
	\setlength{\abovecaptionskip}{-5pt}
	\setlength{\belowcaptionskip}{-10pt}
	\centering
	\begin{minipage}[t]{0.30\linewidth}
		\centering
		\includegraphics[width=2.50in]{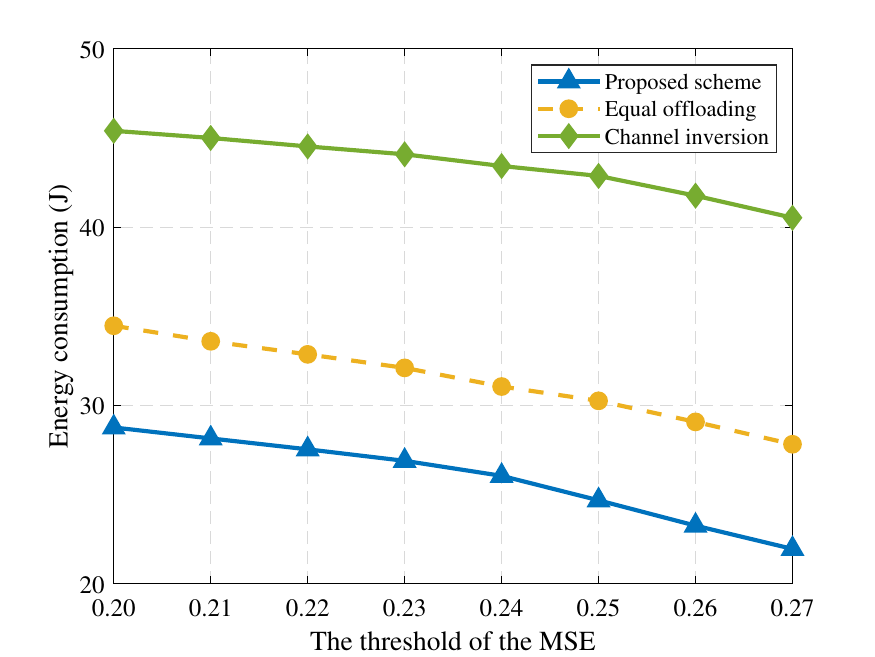}
		\caption{Energy consumption versus the MSE threshold.}
		\label{energy_mse_thresold}
	\end{minipage}%
\hspace{0.5cm}  % 增加水平间距（根据需要调整）
	\begin{minipage}[t]{0.30\linewidth}
		\centering
		\includegraphics[width=2.5in]{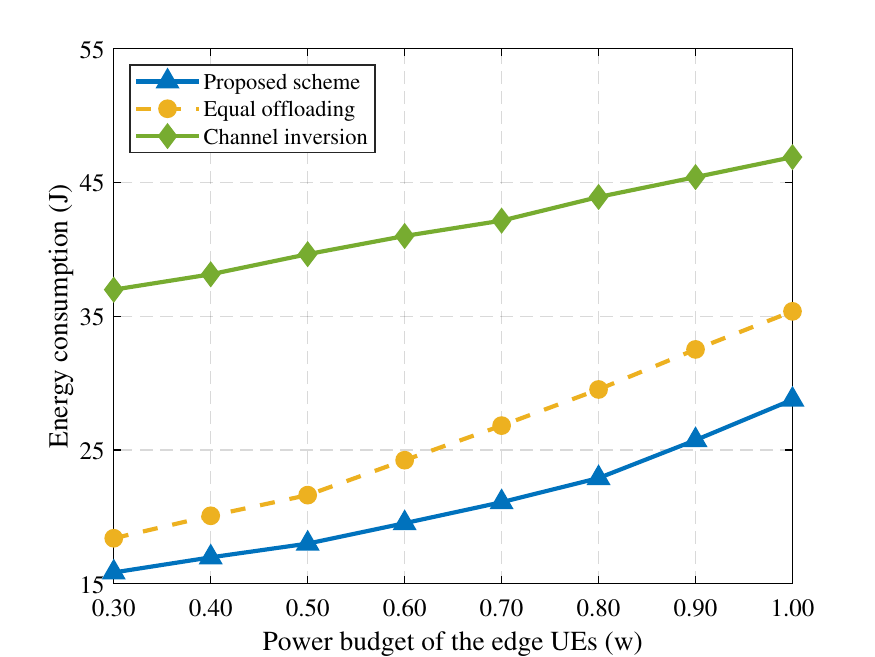}
		\caption{Energy consumption versus the power budget of the edge UEs.}
		\label{energy_power_edge}
	\end{minipage}
\hspace{0.5cm}  % 增加水平间距（根据需要调整）
	\begin{minipage}[t]{0.30\linewidth}
		\centering
		\includegraphics[width=2.5in]{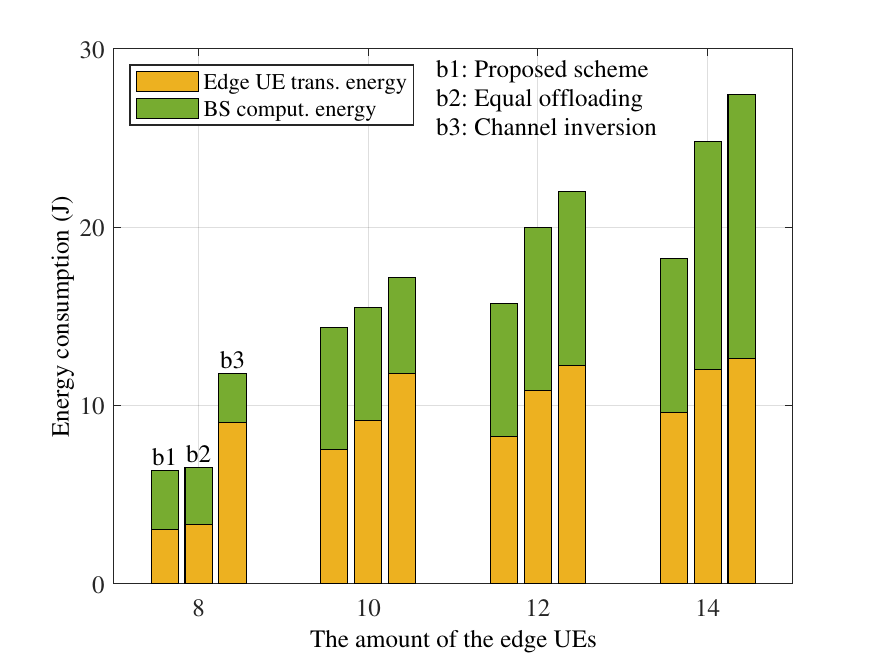}
		\caption{Energy consumption varies with the number of edge UEs.}
		\label{energy_edge_vary}
	\end{minipage}
\end{figure*}

In Fig.~\ref{energy_mse_thresold}, we evaluate the impact of AirComp accuracy on system energy consumption. The result reveals a decreasing trend in energy consumption with larger MSE threshold. This is because a larger MSE threshold means that the system is more tolerant of computational errors, which reduces the transmit power of AirComp UEs and the receiver amplification factor $\eta(i)$. In Fig.~\ref{energy_power_edge},  we can see a clear upward trend in total energy consumption as the power threshold increases. The reason for this is that the larger power constraint expands the feasible region of the resource allocation problem, allowing the system to choose a larger transmission power to satisfy the communication and task offloading constraints. In Fig.~\ref{energy_edge_vary}, we verify the effect of user density on both energy components. Both energy components increase when increasing the number of edge UEs while keeping the total number of UEs in the network constant. The main reason for this is that more edge UEs need to offload a higher data which directly increases the transmission energy and computational energy consumption.

\begin{figure}
	\centering
	\includegraphics[width=0.35\textwidth]{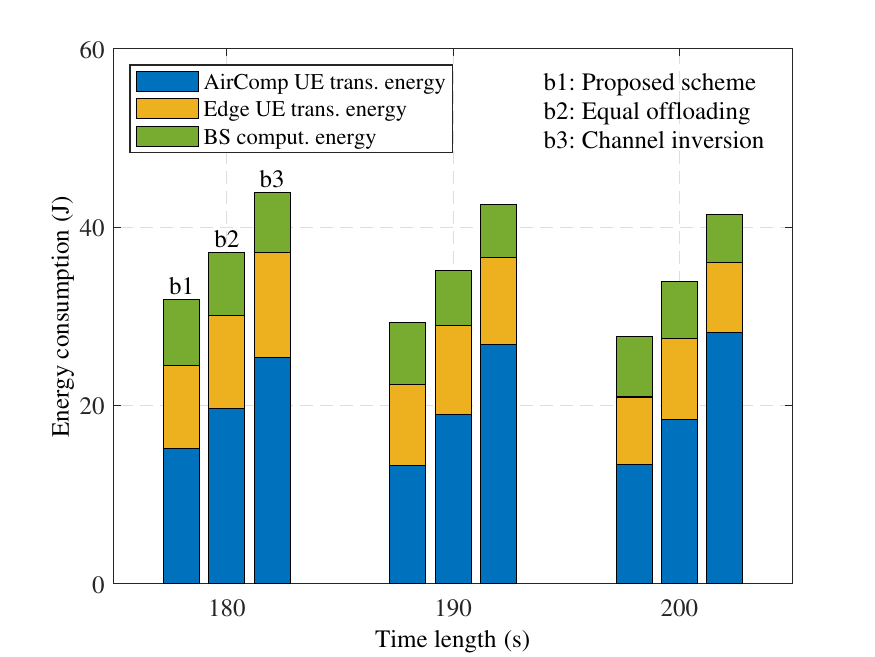}
	\caption{Energy consumption versus the time length.}
	\label{energy_time}
\end{figure}

In Fig.~\ref{energy_time}, we simulate the trend of system energy consumption with time length. We can see that the total system energy consumption degrades gradually with the time length becomes longer. The reason for this is that the longer time length offers the system more freedom for resource optimization. It allows edge computing tasks to be performed more smoother and lower the dependence on AirComp paths and thus reduce the proportion of high energy consumption factors.
\begin{figure}
	\centering
	\includegraphics[width=0.35\textwidth]{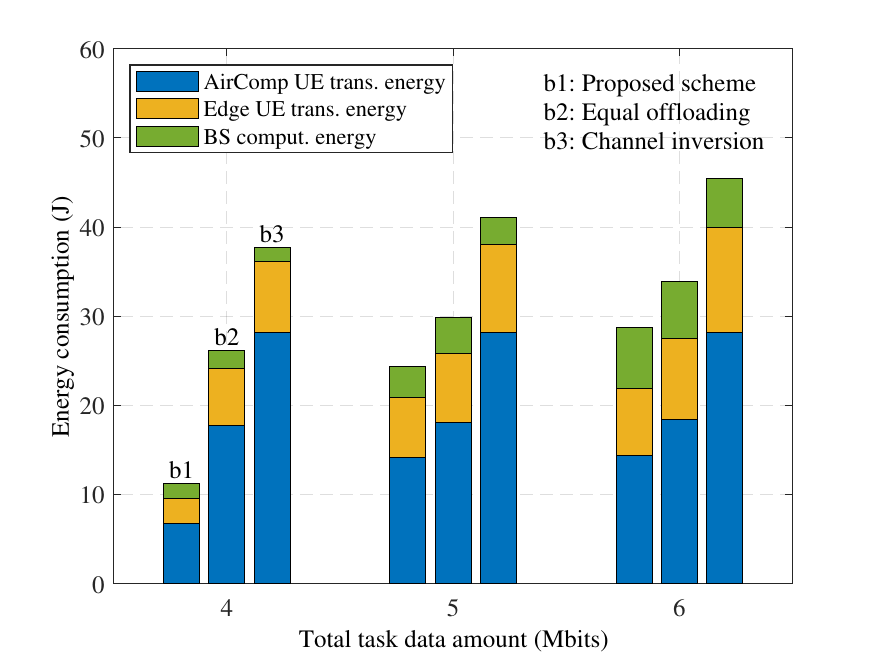}
	\caption{Energy consumption under different offloaded data amount.}
	\label{energy_data}
\end{figure}

In Fig.~\ref{energy_data},  we compare the energy consumption at different offloading data amount. Obviously, the more data is offloaded, the more energy is consumed not only for transmission but also for edge computing. Moreover, the scheme proposed in this paper can lower the energy consumption among all schemes. 

\section{Conclusions}
In this paper, we investigated a hybrid computing system that integrates both edge computing and over-the-air computation (AirComp) to support heterogeneous computing tasks. By formulating a optimization problem under communication, computation, and MSE-based accuracy constraints, we aimed to minimize the system energy consumption. Simulation results validate the effectiveness of the proposed scheme in significantly reducing the system energy consumption.

\bibliographystyle{IEEEtran}
\bibliography{main}

\end{document}